\documentclass[sigconf]{acmart}

\AtBeginDocument{%
  \providecommand\BibTeX{{%
    \normalfont B\kern-0.5em{\scshape i\kern-0.25em b}\kern-0.8em\TeX}}}

\setcopyright{acmcopyright}
\copyrightyear{2018}
\acmYear{2018}
\acmDOI{XXXXXXX.XXXXXXX}

\copyrightyear{2023}
\acmYear{2023}
\setcopyright{acmlicensed}\acmConference[WWW '23]{Proceedings of the ACM Web Conference 2023}{May 1--5, 2023}{Austin, TX, USA}
\acmBooktitle{Proceedings of the ACM Web Conference 2023 (WWW '23), May 1--5, 2023, Austin, TX, USA}
\acmPrice{15.00}
\acmDOI{10.1145/3543507.3583864}
\acmISBN{978-1-4503-9416-1/23/04}

\setcopyright{none} 
\begin{document}

\title{Graph-based Village Level Poverty Identification}


\author{Jing Ma}
\affiliation{%
  \institution{University of Electronic Science and Technology of China}
  \city{Chengdu}
  \country{China}}
\email{jingma@uestc.edu.cn}

\author{Liangwei Yang}
\authornote{Corresponding author}
\affiliation{%
  \institution{University of Illinois Chicago}
  \city{Chicago}
  \country{USA}}
\email{lyang84@uic.edu}

\author{Qiong Feng}
\affiliation{%
  \institution{Southwest Minzu University}
  \city{Chengdu}
  \country{China}}
\email{fengqiongaaron@foxmail.com}

\author{Weizhi Zhang}
\affiliation{%
  \institution{University of Illinois Chicago}
  \city{Chicago}
  \country{USA}}
\email{wzhan42@uic.edu}

\author{Philip S. Yu}
\affiliation{%
  \institution{University of Illinois Chicago}
  \city{Chicago}
  \country{USA}}
\email{psyu@cs.uic.edu}

\renewcommand{\shortauthors}{Trovato and Tobin, et al.}

\begin{abstract}
Poverty status identification is the first obstacle to eradicating poverty. Village-level poverty identification is very challenging due to the arduous field investigation and insufficient information. The development of the Web infrastructure and its modeling tools provides fresh approaches to identifying poor villages. Upon those techniques, we build a village graph for village poverty status identification. By modeling the village connections as a graph through the geographic distance, we show the correlation between village poverty status and its graph topological position and identify two key factors (Centrality, Homophily Decaying effect) for identifying villages. We further propose the first graph-based method to identify poor villages. It includes a global Centrality2Vec module to embed village centrality into the dense vector and a local graph distance convolution module that captures the decaying effect. In this paper, we make the first attempt to interpret and identify village-level poverty from a graph perspective.

\end{abstract}




\maketitle

\section{Introduction}
Ending poverty is the common mission of mankind, which is listed as the first pivotal goal of the United Nations’ Sustainable Development\footnote{https://www.un.org/sustainabledevelopment/poverty/}.
To eradicate extreme poverty for all people everywhere, a fundamental and critical question is where those vulnerable populations are located.
This question refers to general poverty reduction policy interventions to provide material assistance to the poor, such as ``where should the next school or main road be?''~\cite{bedi2007more,lee2022high} As the basic socio-economic unit, the village is always seen as the cell of the social system~\cite{Li2018What}; so poverty identification at the village level has become the key to mapping poverty.

The household surveys and population census are the standard ways to measure an area/individual's socioeconomic conditions, which provide policymakers with critical statistics for mapping out resource assignments~\cite{Charlotte2018Earth,guo2022estimating,ma2022counterfactual}. However, with the rapid socioeconomic and demographic changes, data collection at a higher frequency is required, which means substantial costs~\cite{njuguna2017constructing}. Besides, due to the diverse sources of income and the asymmetric information between the investigators and interviewees, the reliability and validity of survey data are doubted. The development of Web infrastructure and modeling tools provide new opportunities and fresh approaches to identify poor villages. In recent years, combining geospatial information and machine learning technology has become ever increasing interest for research on poverty 
area identification\cite{hargreaves2021satellite,guo2016big,jean2016combining}. 
Geospatial information, such as nighttime lights, day-time satellite imagery, and crowd-sourced map data, can assist in capturing poverty and socioeconomic conditions on a coarse scale\cite{2020Utilizing,burke2021using,xu2021combining}. Machine learning technology allows researchers to effectively and efficiently utilize geospatial information\cite{hu2022village,hersh2021open,wide}. 
However, these methods rely too much on obtaining quantifiable geospatial features while some information, such as nighttime light, is unfeasible to collect at the village level. Besides, these methods pay much attention to geographical characteristics but ignore the relationship between villages that shows regional economic activities.



In this study, we propose a novel method to identify poor villages based on the Web infrastructure and its widely applied graph-based modeling methods~\cite{yang2022large,dou2022robust,schinas2015multimodal,yang2021consisrec}. We
build the village graph based on distances and analyze the poverty occurrence from the graph perspective. Through field investigation, we collected village poverty labels in Enshi prefecture, one of the poverty-prone cities in China, and obtained the geological location by web map services. We identify two factors in the village graph to model poverty occurrence. 1) Village centrality in the graph, 2) Village's distance homophily decay effect. Based on the observation, we designed a graph-based model to identify poor villages. A global Centrality2Vec module to capture the centrality similarity between nodes. 
We reconstruct the edges based on different kinds of nodes' centrality measures and perform random-walk-based skip-gram training~\cite{word2vec} to obtain centrality-aware node features.
A local graph distance convolution is designed to aggregate information from direct neighborhoods, where we model the homophily decay effect as the decayed edge weight based on distance. The collected data and code are open-sourced at \textcolor{blue}{\url{https://github.com/YangLiangwei/Graph-Poverty-Identification}}. Our contributions are summarized as follows:
\begin{itemize}
    \item We conducted a field investigation to collect village-level poverty data and released it to the research community.
    \item We make the first attempt to analyze and identify village poverty occurrence from the graph perspective.
    \item A graph-based model is designed accordingly to identify poor villages from the geologic topology.
\end{itemize}

\section{Poverty Data Analysis}
In this section, we present the study area, data collection procedure, and related data analysis.

\begin{figure}
    \begin{center}
    \includegraphics[width=0.475\textwidth]{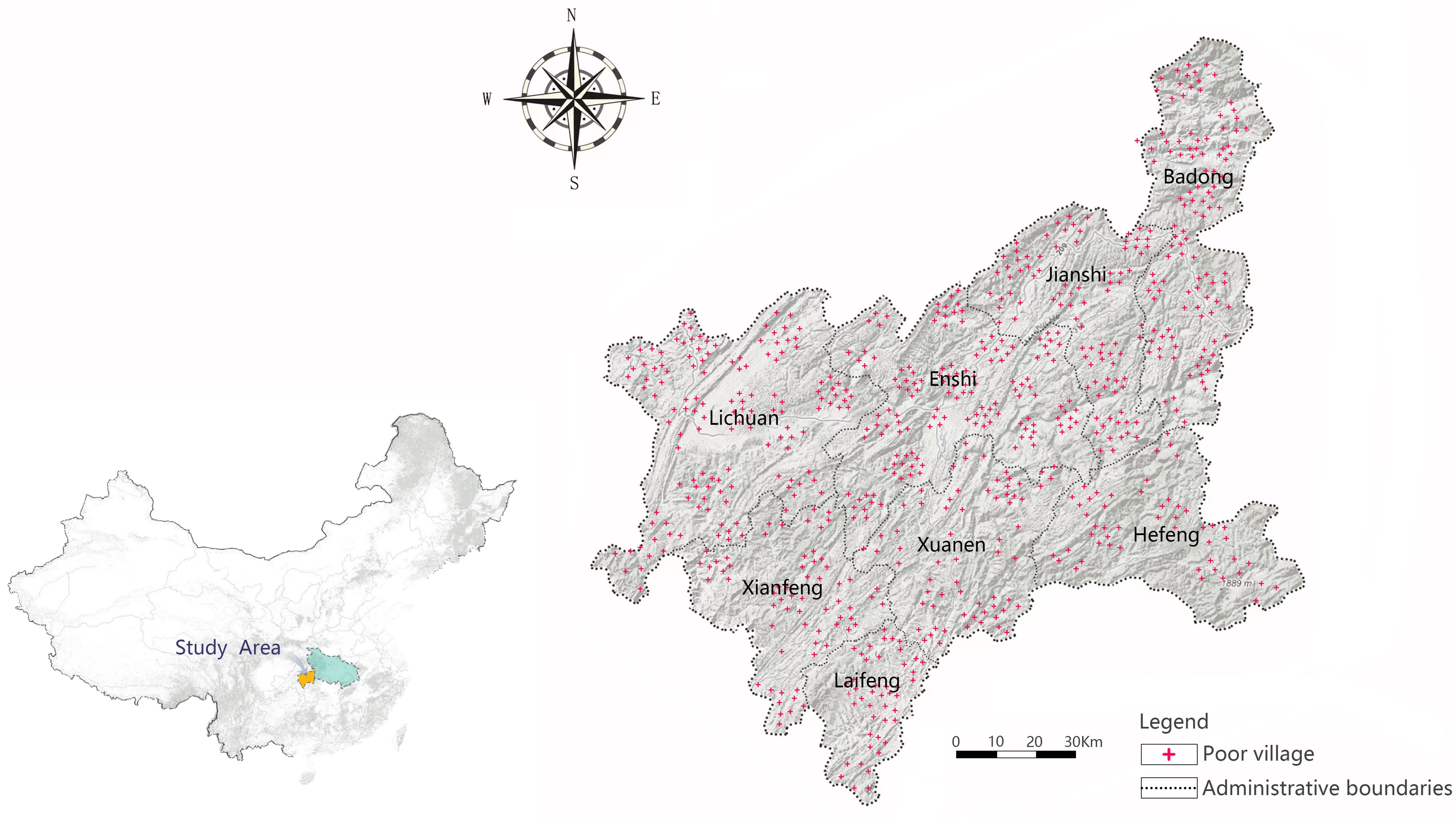}
    \end{center}
    \caption{Study Area}
    \label{fig:area}
\end{figure}

\subsection{Study Area}
The study area, Enshi, is located in the southwest corner of Hubei Province, China (shown in Fig.~\ref{fig:area}). This area covers 24,060.26 $km^2$, with a distance of about 220 kilometers ranging from east to west and the distance of about 260 kilometers ranging from north to south. As a profoundly impoverished area, Enshi has 3.46 million permanent residents, with 1.09 million being poor. All residents are clustered in 2,606 villages. In 2013, the first year that China started the target poverty alleviation (TPA) program~\cite{Zhou2018Targeted,hu2022detection,qiong2021Poverty}, 729 among 2,606 villages were identified as being poor.
As a poor region, the per capita GDP of Enshi in 2021 is 34,300 CNY, which is significantly lower than the national figure of 70,900 CNY; the per capita disposable income of these rural residents is 11,600 CNY, which is significantly lower than the national figure of 16,000 CNY. It is typical to take Enshi Prefecture as an example to study the village poverty problem of mountainous areas.

\subsection{Data Collection}
We surveyed through face-to-face interviews with County Poverty Alleviation Office leaders using a semi-structured questionnaire in Enshi. 
The data covers the number of people in each administrative village, the number of poor people, and the incidence of poverty.

The village graph $\mathcal{G_{\text{village}}}=\{\mathcal{V}, \mathcal{E}\}$ is built to represent the connections between villages, where $\mathcal{V}=\{v_1, v_2,\cdots,v_n\}$ is the village set and $\mathcal{E}=\{e_1, e_2,\cdots,e_m\}$ is the edge set. The edge $e = (v_i,v_j)$ is constructed if the distance between $v_i$ and $v_j$ is smaller than a distance threshold $d$.
The distance can reflect their geographical relationship. The economic relationship is also reflected because a closer distance usually indicates more frequent economic exchanges. Taking advantage of online web tools, we utilize the Google map\footnote{https://developers.google.com/maps} to acquire the latitude and longitude of each village as $v_i = (lat_i, lon_i)$. Then we calculate the geodesic distance~\cite{karney2013algorithms} between two villages on the surface of the ellipsoidal model of the Earth. One edge forms if the distance between two villages is smaller than $d$.

\subsection{Node Centrality Analysis}\label{sec:centrality}
\begin{figure}
    \begin{center}
    \includegraphics[width=.22\textwidth]{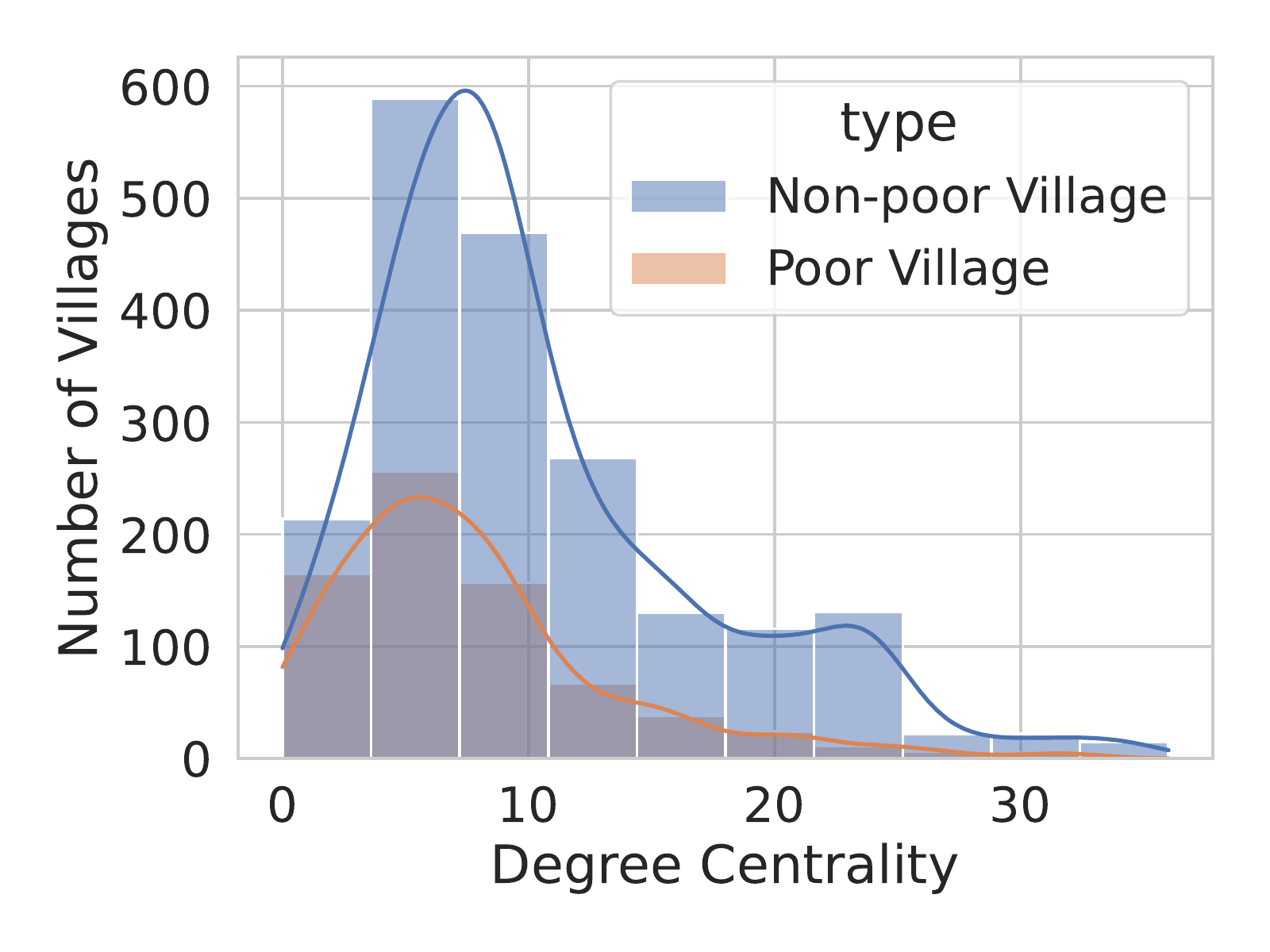}
    \includegraphics[width=.22\textwidth]{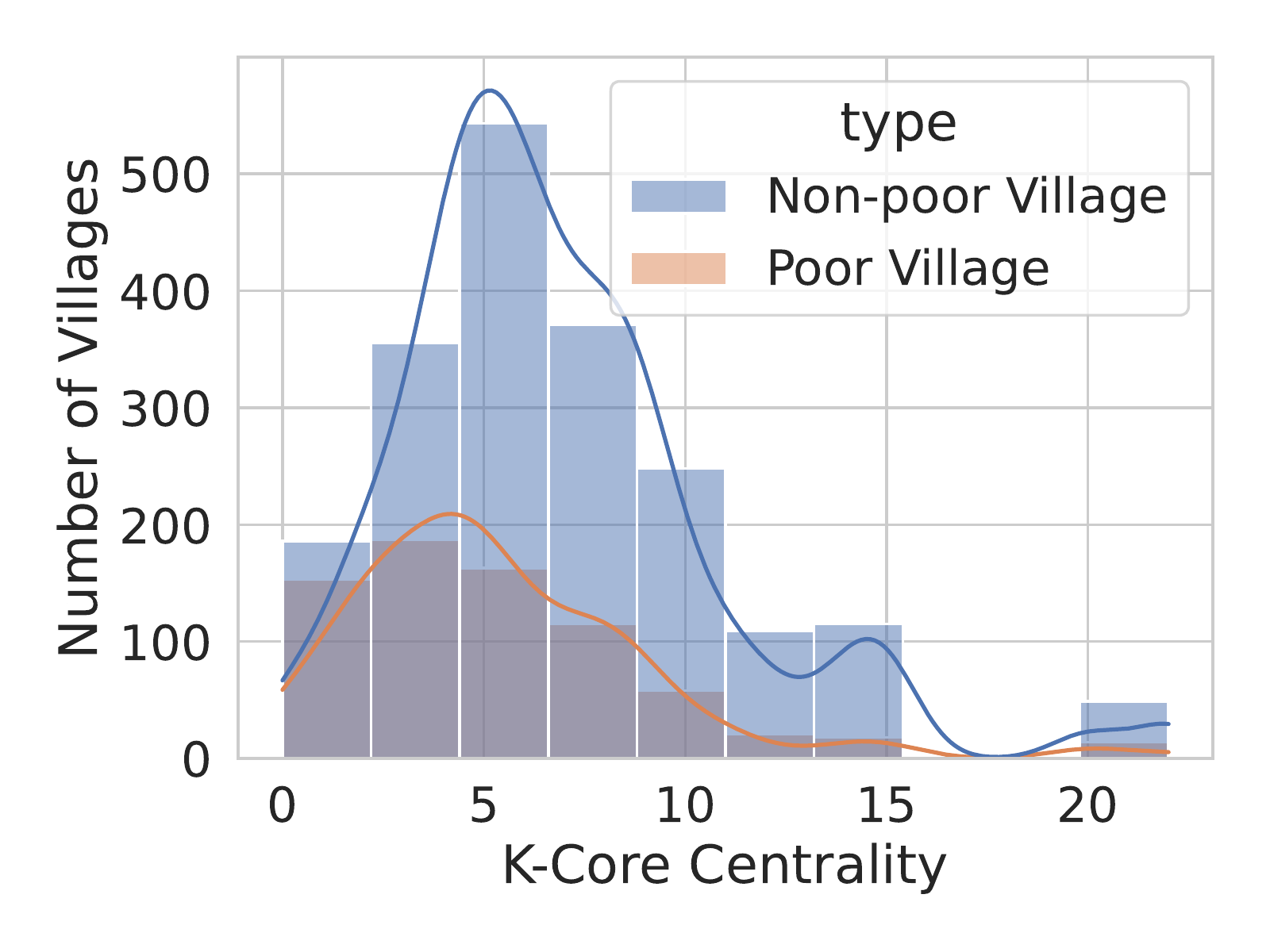}
    \end{center}
    \caption{Node Centrality analysis on Villages}
    \label{fig:centrality}
\end{figure}

A place that has rich natural resources and a suitable living environment tends to gather households and form a village cluster. Simultaneously, economic activities are also boosted. The clustering effect can be reflected by the node centrality measure on village graph $\mathcal{G_{\text{village}}}$. Thus, we first analyze the relationship between village node centrality and poverty occurrence. The results of $t$ test show that the centrality of the poor villages is smaller than that of the non-poor ($p<0.001$). The poor/non-poor village distribution concerning two kinds of centrality measures, degree, and K-Core~\cite{kcore} also illustrate that (Fig.~\ref{fig:centrality}). The degree centrality distribution for poor villages is $7.60 \pm 5.68$ while for non-poor villages is $10.31 \pm 6.82$. The K-Core centrality distribution for poor villages is $5.41 \pm 3.76$ while for non-poor villages is $6.88 \pm 4.01$. Node centrality can reveal the village's poverty status to some extent. Nodes with similar centrality are not necessarily connected. Global structural information is required to capture centrality similarity for village-level poverty identification. It motivates the design of Global Centrality2Vec.

\subsection{Distance Analysis}\label{sec:distance}
\begin{figure}
    \begin{center}
    \includegraphics[width=.23\textwidth]{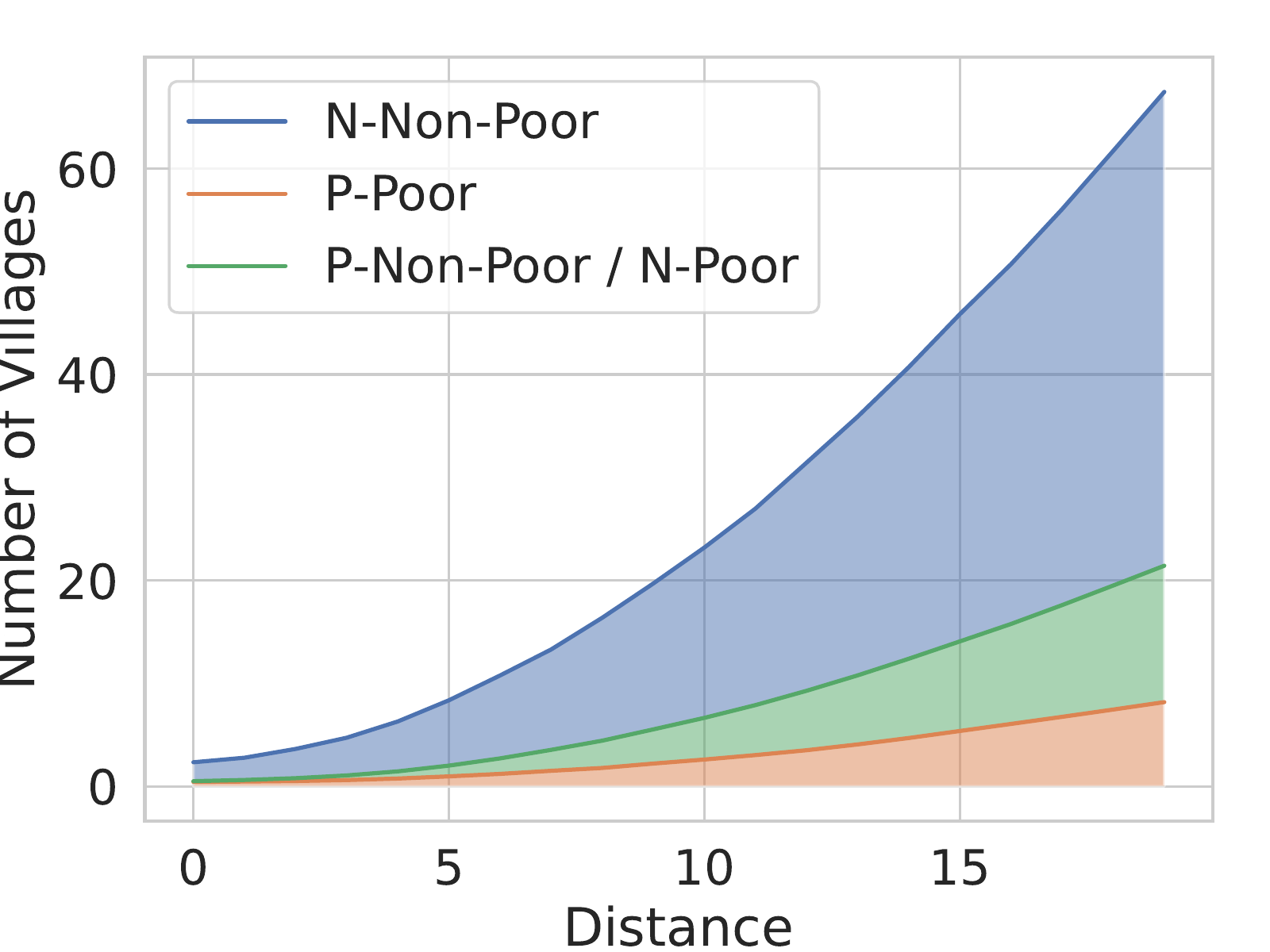}
    \includegraphics[width=.23\textwidth]{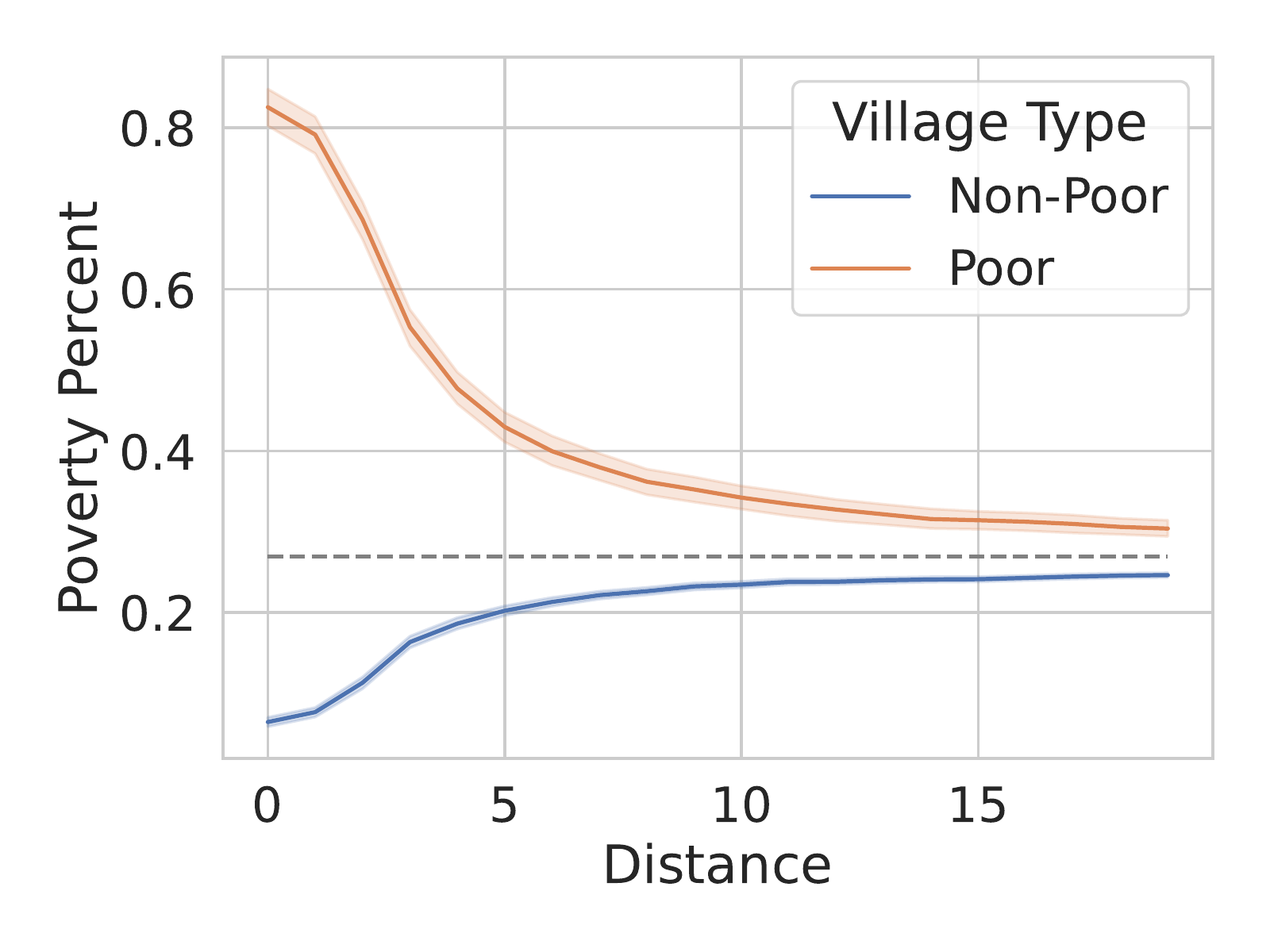}
    \end{center}
    \caption{Distance analysis on Villages}
    \label{fig:distance}
\end{figure}

Distance between villages can reflect the transportation difficulty, which has a direct influence on the economic activity between villages.
In Fig.~\ref{fig:distance}(left), we show the increase of different types of neighborhoods when gradually increasing the distance. The ``P Poor'' indicates the number of poor village neighborhood of the poor center village, ``P Non Poor'' represents the number of non-poor villages in the neighborhood of the poor center village, etc. 
In Fig.~\ref{fig:distance} (right), we present the change in poor village percent with the increase of village distance. With the increase in distance, the poor village percentage in neighborhoods is decreasing for poor villages while it is increasing for non-poor villages. It shows the homophily effect~\cite{homophily} on $\mathcal{G_{\text{village}}}$ that geologically near villages tend to have similar poverty status. The homophily effect is within a local scope, which decreases with the increase in distance. Based on the analysis, we model the homophily decay effect into message passing and propose the graph distance convolution to aggregate local neighborhood information.

\section{Method}
\begin{figure*}
    \begin{center}
    \includegraphics[width=\textwidth]{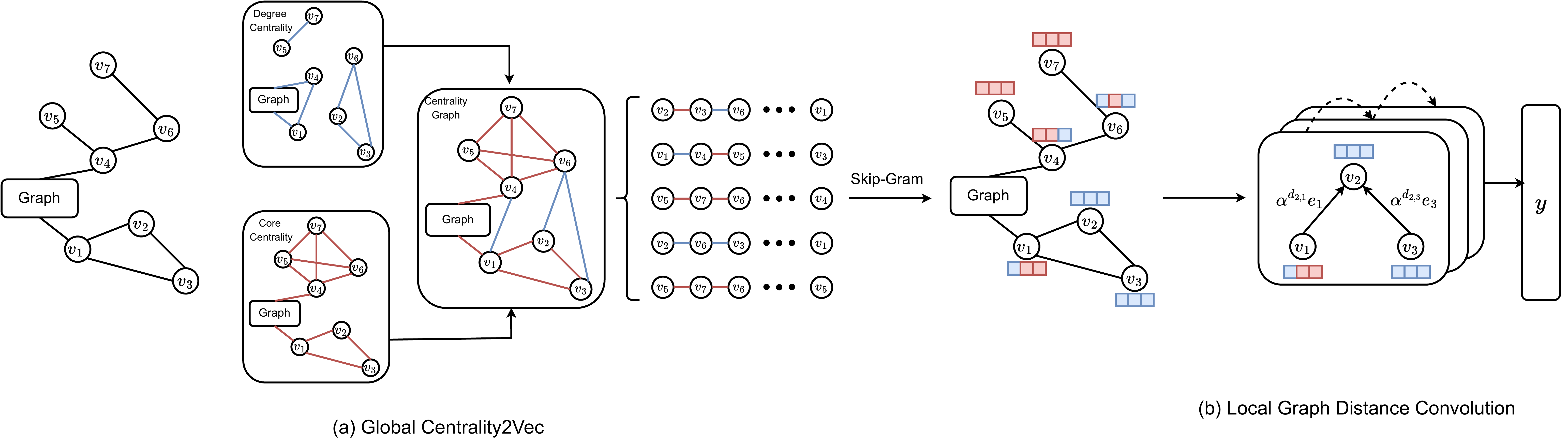}
    \end{center}
    \caption{Model Framework}
    \label{fig:framework}
\end{figure*}

In this section, we introduce the designed graph-based model. As shown in Fig.~\ref{fig:framework}, it consists of a global Centrality2Vec module to capture the village's centrality similarity, and a local graph distance convolution module that aggregates neighborhood information.

\subsection{Global Centrality2Vec}\label{sec:global}
The analysis from Sec.~\ref{sec:centrality} shows poor villages tend to have a smaller centrality in $\mathcal{G_\text{village}}$. We propose Centrality2Vec to embed different kinds of centrality into one dense vector. 
Centrality2Vec first computes different kinds of centrality for each node such as degree and core centrality. To capture the surrounding structure topology as in struct2vec~\cite{struct2vec}, we compute village centrality similarity based on the ordered centrality sequences of villages within $1-$hop in $\mathcal{G_\text{village}}$. The ordered centrality sequences of village $v_i$ and $v_j$ are represented as $S_i=(c_1^i,c_2^i,\cdots,c_n^i)$ and $S_j=(c_1^j,c_2^j,\cdots,c_m^j)$, respectively. As the number of neighborhoods is not the same for each node, the length of $S_i$ and $S_j$ can be different. To measure the similarity between two different length sequences, we compute the pair-wise village similarity based on Dynamic Time Warping (DTW)~\cite{dtw}. DTW computes the change cost from $S_i$ to $S_j$ as:
\begin{equation}
    \text{Cost}(S_i, S_j)=\sum_{c^j\in S_j}\min_{c^i \in S_i}\text{cost}(c^j,c^i).
\end{equation}
For pair of element $c^j$ and $c^i$, the transform cost is defined as:
\begin{equation}
\text{cost}(c^j,c^i)=\frac{max(c^j,c^i)}{min(c^j,c^i)}-1.
\end{equation}
For each village, we add edges to its most similar centrality villages with the Top-K smallest cost. For each centrality measure, we obtain one centrality similarity graph. In this paper, we utilize degree and core centrality to obtain $\mathcal{G_\text{degree}}$ and $\mathcal{G_\text{core}}$, respectively. To capture the centrality similarity of different measures, we combine the two graphs and define the random-walk transition probability as:
\begin{equation}
    p(v_j|v_i)=\begin{cases}
    \frac{1}{|\mathcal{N}_{\text{degree}}^i + \mathcal{N}_{\text{core}}^i|} & (v_i,v_j) \in (\mathcal{E}_{\text{degree}} \cup \mathcal{E}_{\text{core}}) \\
    0 & \text{Otherwise},
    \end{cases}
\end{equation}
where $\mathcal{N}_{\text{degree}}^i$ is the neighbor set of $v_i$ in $\mathcal{G}_\text{degree}$, and $\mathcal{E}_{\text{degree}}$ is the edge set of $\mathcal{G}_\text{degree}$. The definition is the same for $\mathcal{G}_\text{core}$. Then we perform random walks on the combined transition probability matrix to collect walk sequences $W_k=(v_1^k, v_2^k, \cdots, v_n^k)$, where $v_i \in \mathcal{V}_\text{village}$. Based on the collected sequences, we aim to learn village representation $\mathbf{H}\in \mathbb{R}^{|\mathcal{V}|*d}$ that can capture the centrality similarity, where $v_i$ is represented as $h_i$. Then we use the skip-gram model to update $\mathbf{H}$ by optimizing the context occurrence loss as:
\begin{eqnarray}
    \mathcal{L}_{\mathbf{H}}&=&-\sum_{W}\log P(\{v_{i-z}^W,\cdots,v_{i+z}^W \}/ v_i^W|v_i^W) \\
    &=&-\sum_{W}\prod_{j=i-z,j\neq i}^{i+z}\log P(v_j^W|v_i^W),
\end{eqnarray}
where $W$ is the random walked village sequences, $z$ is the window size for training, and the probability $P(v_j|v_i)$ is calculated by:
\begin{equation}
    P(v_j|v_i)=\frac{\text{exp}(\mathbf{h}_j\cdot \mathbf{h}_i)}{\sum_{e \in \mathbf{H}}\text{exp}(\mathbf{h}\cdot \mathbf{h}_i)}.
\end{equation}
Compared with struct2vec, Centrality2Vec focuses more on node centrality. It explicitly utilizes multiple kinds of node centrality measures. For village $v_i$, it embeds different kinds of centrality similarity into one dense vector $\mathbf{h}_i$, which is used as the input feature to the local graph distance convolution module.

\subsection{Local Graph Distance Convolution (LGDC)}\label{sec:local}
Distance analysis from Sec~\ref{sec:distance} shows the homophily effect on $\mathcal{G_{\text{village}}}$. We design a local graph convolution module to aggregate neighborhood information. A general graph neural network (GNN) can be formalized as:
\begin{equation}
    \mathbf{h}_{i}^{(l+1)} = \mathbf{h}_{i}^{(l)} \oplus \text{AGG}^{(l+1)} (\{ \mathbf{h}_{j}^{(l)} \mid v_j \in \mathcal{N}_i \}),
\end{equation}
where $\mathbf{h}_{i}^{(l)}$ is $v_i$'s embedding on $l$-th layer, $\mathcal{N}_i$ is the neighbor set of $v_i$, AGG is the aggregation function on neighborhood information, and $\oplus$ is the reduction function to combine the neighborhood information and node's own embedding. 
We model the distance-based homophily decaying effect into the aggregation function as:
\begin{equation}
    \text{AGG}^{(l+1)} (\{ \mathbf{h}_{j}^{(l)} \mid v_j \in \mathcal{N}_i \})=\sum_{v_j\in \mathcal{N}_i}\frac{\alpha^{\text{dist}_{i,j}}}{\sqrt{|\mathcal{N}_i|}\sqrt{|\mathcal{N}_j|}}\mathbf{h}_j,
\end{equation}
where $\text{dist}_{i,j}$ is the distance between $v_i$ and $v_j$, $0\le \alpha \leq 1$ is the hyper-parameter to control the decaying effect. When $\alpha = 1$ indicates there is no decaying. The reduction function is defined as:
\begin{equation}
    e_i^{(l)} \oplus \text{AGG}^{(l+1)}(\cdot)=(\frac{1}{|\mathcal{N}_i|}\mathbf{h}_i + \text{AGG}^{(l+1)}(\cdot))\mathbf{W} + b,
\end{equation}
where $\mathbf{W}$ and $b$ is the learnable parameters for graph distance convolution. After several layers of graph convolution, we map the village to a $2$ dimensional vector to predict the village type. ReLU activation function is applied between layers.

\section{Experiments}
In this section, we conduct experiments to test the model's effectiveness and the influence of designed modules.

\subsection{Experimental Setup}

\begin{table}
  \caption{Statistics of the Collected Datasets}
  \label{data stat}
  \scalebox{0.9}{
  \begin{tabular}{l c}
        \toprule
        Type & Number \\
        \midrule
        Poor Villages & 729 \\
        Vulnerable Population & 1,104,931\\
        Non-poor Villages & 1,976 \\
        Non-vulnerable Population & 2,439,925 \\
        \hline
        Graph Edges ($d=5km$) & 28,613 \\
        Node Average Degree ($d=5km$) & 10.58 \\
        Graph Sparsity & $0.3910\%$ \\
        \bottomrule
  \end{tabular}}
\end{table}

The dataset statistics is shown in Table~\ref{data stat}. The collected dataset covers 2,705 villages with more than 3 million population. We construct village graph $\mathcal{G_\text{village}}$ with a distance threshold $d=5km$. 
We compare our model with two kinds of baselines. Struct2Vec~\cite{struct2vec} is a graph embedding method that can capture node structure similarity. The other kind is GNN models including GCN~\cite{gcn}, GAT~\cite{gat}, SGC~\cite{sgc} and APPNP~\cite{appnp}. For a fair comparison, we perform the same grid search and keep the layers as $2$ for all models.

\subsection{Performance Evaluation}

\begin{table}[htbp]
    \caption{Overall comparison, the best and second-best results are in bold and underlined, respectively}
    \label{tab:comparison}
    \centering
    \scalebox{0.9}{
    \begin{tabular}{l|ccccc}
         \hline
         Model & Accuracy & Precision & Recall & F1 & AUROC \\
         \hline 
         Struct2Vec & 0.7282 & 0.5723 & 0.5326 & \underline{0.5214} & 0.5859 \\
         GCN & 0.7449 & 0.6929 & 0.5385 & 0.5090 & 0.5642 \\
         GAT & 0.7338 & 0.6214 & 0.5221 & 0.4825 & 0.5302 \\
         SGC & 0.7301 & 0.4918 & 0.4996 & 0.4285 & 0.5070 \\
         APPNP & \underline{0.7504} & \underline{0.7573} & \underline{0.5401} & 0.5074 & \underline{0.5880} \\
         Our method & \textbf{0.7607} & \textbf{0.7926} & \textbf{0.5612} & \textbf{0.5434} & \textbf{0.6164} \\
         \hline
    \end{tabular}}
\end{table}

Experiment results are shown in Table~\ref{tab:comparison}. We can observe that our model achieves the best performance on all metrics, which indicates our model can effectively utilize the global centrality similarity and the local neighborhood information. Compared with accuracy, the recall score is much less. It is because of the label imbalance. Only about $1/4$ of the villages are poor-villages, which makes all models tend to predict villages as non-poor. At last, though we set the first baseline for graph-based village identification, the task is still challenging and there is still much room for improvement.

\subsection{Model Analysis}

\begin{figure}
    \begin{center}
    \includegraphics[width=.2\textwidth]{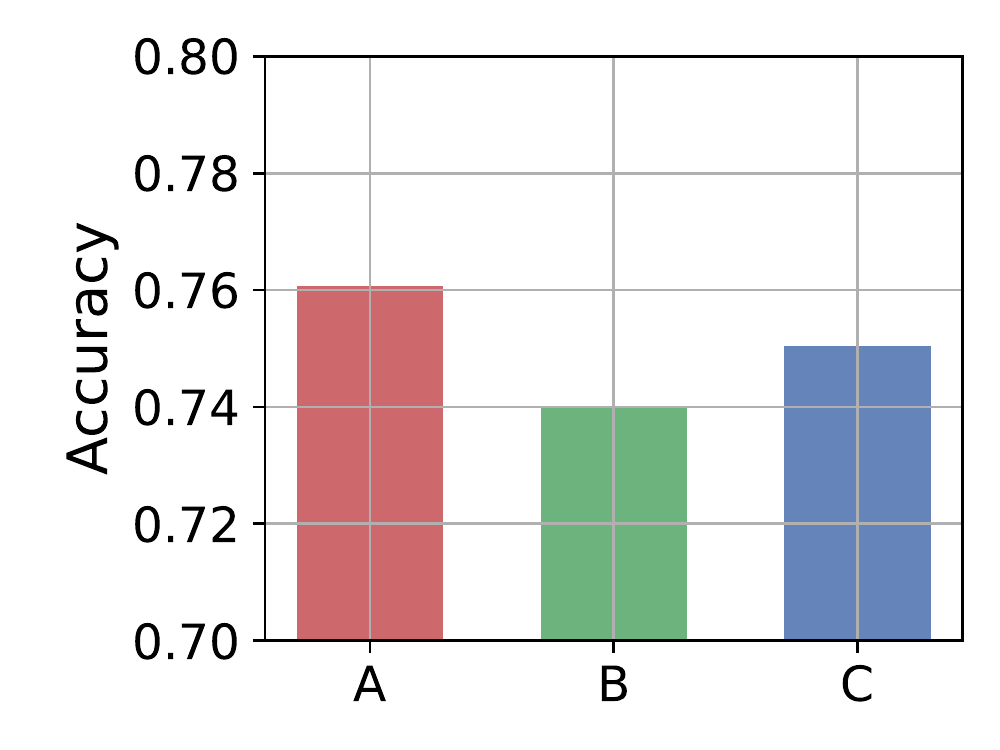}
    \includegraphics[width=.2\textwidth]{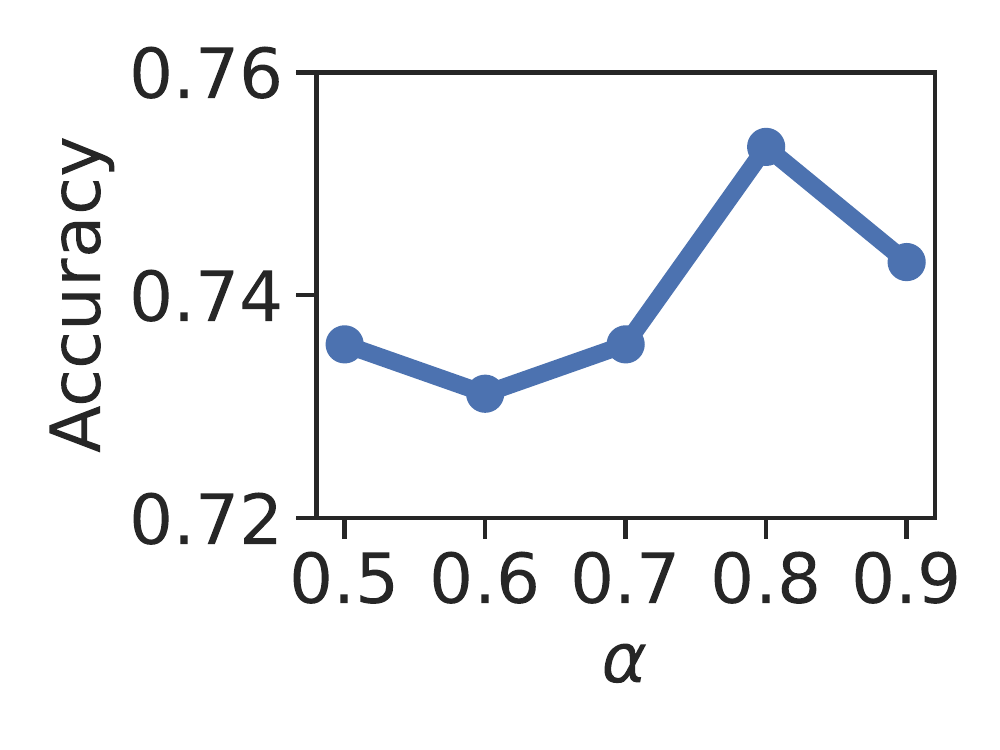}
    \end{center}
    \caption{Model analysis. Left: ablation study, where A is our model, B is w/o Global Centrality2Vec, C is w/o local graph distance convolution. Right: the impact of $\alpha$.}
    \label{fig:analysis}
\end{figure}

We make further model analysis based on the ablation study and the sensitivity of $\alpha$ designed in the local graph distance convolution module. The ablation study is shown in Fig.~\ref{fig:analysis} (left). We can observe that the whole model achieves the best. It shows the joint modeling of global centrality similarity and local graph distance convolution is effective. The model w/o Centrality2Vec performs the worst, which indicates the importance of village centrality similarity. We then show the influence of $\alpha$ in Fig.~\ref{fig:analysis} (right). $\alpha$ decides the distance decaying effect during graph convolution. Larger $\alpha$ indicates slower decaying. With the increase of $\alpha$, accuracy first increases to its peak at $\alpha =0.8$ then drops. It shows a suitable decaying speed that fits the data can boost performance, which also validates the effectiveness of modeling the decaying effect into local graph convolution.

\section{Conclusion}
Poverty is still a challenging problem faced by mankind.
In this paper, we make the first attempt to identify village-level poverty status from the graph perspective. 
By connecting villages as a graph based on geographic distance, we observe two key factors (Centrality, Homophily Decay effect) for identification.
Accordingly, we design a global Centrality2Vec and a local graph distance convolution module to identify poor villages. We further open-sourced the collected poverty data to the community for further research.


\begin{acks}
This work is supported in part by NSF under grants III-1763325, III-1909323,  III-2106758, SaTC-1930941, and the "Fundamental Research Funds for the Central Universities”, Southwest Minzu University (2023SQN11).
\end{acks}

\bibliographystyle{ACM-Reference-Format}
\bibliography{sample-base}

\appendix

\end{document}